# A two-step Recommendation Algorithm via Iterative Local Least Squares

Jinhu Liu[1], Chengcheng Yang[1], Zi-Ke Zhang[2, 1]

1. Web Sciences Center, University of Electronic Science and Technology of China, Chengdu, 611731, P.R. China
2. Institute of Information Economy, Hangzhou Normal University, Hangzhou, 310036, P.R. China
{ljinhu,yangch905,zhangzike}@gmail.com

**Abstract.** Recommender systems can change our life a lot and help us select suitable and favorite items much more conveniently and easily. As a consequence, various kinds of algorithms have been proposed in last few years to improve the performance. However, all of them face one critical problem: data sparsity. In this paper, we proposed a two-step recommendation algorithm via iterative local least squares (ILLS). Firstly, we obtain the *ratings* matrix which is constructed via users' behavioral records, and it is normally very sparse. Secondly, we preprocess the "ratings" matrix through ProbS which can convert the sparse data to a dense one. Then we use ILLS to estimate those missing values. Finally, the recommendation list is generated. Experimental results on the three datasets: *MovieLens, Netflix, RYM*, suggest that the proposed method can enhance the algorithmic accuracy of AUC. Especially, it performs much better in dense datasets. Furthermore, since this methods can improve those missing value more accurately via iteration which might show light in discovering those inactive users' purchasing intention and eventually solving *cold-start* problem.

## 1 Introduction

The last few years have witnessed the rapid development of Internet and social networks, we are confronting too much information on the World Wide Web and have to spend too much time to find which is more relevant for us [1,2]. Especially in recent years, Web 2.0 techniques have facilitated the emergence of a plenty of websites enabling large-scale communities to aggregate and interact [3]. For example, *Facebook*, launched in February 2004, has attracted more than 900 million active users within two years. Amazon has already reported over 1.3 million sellers selling products through Amazon's World Wild Web sites in 2007[1] . In April 2011, Netflix announced 23.6 million subscribers in the United States and over 26 million world-wide[2]. Therefore, all those platforms, as well as many other applications, have provided versatile sources to explore in filtering out redundant and irrelevant information.

---

[1] http://en.wikipedia.org/wiki/Amazon.com

Actually, in 1992, David Goldberg *et al.* have developed the first filtering system [4], which allowed users to annotate documents to navigate resources. Subsequently, similar concepts were applied to *Usenet* news by the *Grouplens* research project [5]. In 1999, Schafer *et al.* presented an explanation of how recommender systems help E-commerce sites increase sales, and analyzed six sites that used recommender systems, including several sites that adopted more than one recommender systems. Finally, they created taxonomy of recommender systems based on those methods [6].

Those results gave fresh impetus to the research and development of recommender systems, as well as the widespread commercial applications. Recently, various kinds of algorithms have been proposed, including collaborative filtering (CF) approaches [6-10],content-based analyses[11-12], tag-aware algorithms [13-15], link prediction approaches[16-18], hybrid algorithms [19-20] and so on. And all of them tried to improve the performance of traditional algorithms. Although the emergence of those recommender systems changes our life a lot [21-22], there is still one critical problem: *Data Sparsity*.

Nowadays, there is a vast class of methods proposed from various fields, including mathematics, computer science, complex network, Physics, Biology and so on. For example, Matrix factorization based on Principle Component Analysis or Singular Value Decomposition(SVD) [23-25], personal recommendation via bipartite network projection [16,33] and Heat-Spreading algorithm [20].

In the paper, we proposed a two-step recommendation algorithm by the classical theory in microarray missing values, so-called *Iterative Local Least Squares* (ILLS) [26] which is the iterative application of Local Least Squares (LLS) method [27]. As one of three main methods in microarray missing value imputation, LLS [27] was firstly presented in 2004 by Kim *et al.* They utilized Least Squares method to estimate missing values in DNA microarray. The major difference between LLS and ILLS is whether the number of neighbors is fixed. As the neighbors' number for target gene in ILLS is not fixed but determined by similarity with the target gene, the treatment effect is much better.

The rest of this paper is as following: the related work about ILLS is described in section 2. In section 3, the new recommendation method will be presented. The experiments and results are shown in section 4. Finally, we conclude our work in the last section.

## 2    Related works

For recommendation algorithms, there are already several pioneering works based on iterative concept which is used for improving the quality of items and the reputation of users in order to help customers make right decisions [28-29]. Those models have both successfully provided meaningful rating counts and enhanced the system robustness. Although ILLS has not been applied in recommender systems, ILLS has been used in microarray missing value imputation and obtained better results [26]. Both the microarray missing value imputation and the prediction of items' rating have encountered similar problems. The same or homologous methods always effectively solve them. We have obtained improved results utilizing this concept [6-10] and an object-oriented algorithm, ProbS [19].

---

[2] http://en.wikipedia.org/wiki/Netflix

## 3  Iterative Local Least Squares

### 3.1  Construct Rating Matrix

A recommender system can be represented by a graphic form: $G(U, O, E)$, where $U = \{u_1, u_2, \ldots, u_m\}$, $O = \{o_1, o_2, \ldots, o_n\}$, and $E = \{e_1, e_2, \ldots, e_q\}$ are respectively the sets of users, items and links, respectively [17]. These links can be represented by a $n \times m$ adjacency matrix A where $a_{\alpha i} = 1$ (we use Greek and Latin letters, respectively, for item- and user-related indices) if item α is collected by user i (or i-th user) and $a_{\alpha i} = 0$ otherwise [20]. The set E is used to represent the purchasing or other operating behavior of users. In real life, many online e-commerce websites provide rating function. So the rating matrix can be presented by $R_{n \times m} = (r_{\alpha i})_{n \times m}$ where $r_{\alpha i}$ records the user i's rating to item α. If the user i has not selected the item α, $r_{\alpha i} = 0$. A row $r_i^T (1 \leq i \leq m) \in R^{n \times 1}$ represents the $i - th$ user's rating to all items.
$$R = (r_1^T \quad r_2^T \quad \cdots \quad r_{m-1}^T \quad r_m^T) \in \mathbb{R}^{n \times m}$$
From the data set we have (see section 4), it is certain that rating matrix R is a sparse matrix, which would inference the expected performance. In order to overcome such sparsity problem, we use following method to preprocess above rating matrix.

### 3.2  Preprocess Rating Matrix

In this paper, we use a two-step method to overcome the sparse data problem. First, we adopt the so-called probabilistic spreading (ProbS) algorithm [19] to convert the sparse data to a dense one. Here, R is no more the rating matrix (For convenience, the vectors in R is still called "ratings"), but a binary matrix, which is similar to adjacency matrix A. First, we assign the items that target users have already selected an initial level of "resource" denoted by the vector f (where $f_\beta$ is the resource possessed by itemβ. In our method, we make initial "resource" set to "1"), then redistribute it in a transformation from: $\tilde{f} = W^p f$, where

$$W_{\alpha\beta}^p = \frac{1}{k_\beta} \sum_{j=1}^{m} \frac{a_{\alpha j} a_{\beta j}}{k_j}, \tag{1}$$

is a row-normalized $n \times n$ matrix, $k_\beta$ ($k_j$) is the degree of item β (user j). Then the new definition for R is as following:

$$R = (r_1^T \quad r_2^T \quad \cdots \quad r_{m-1}^T \quad r_m^T) = \begin{pmatrix} r_{11} \cdots r_{11} \cdots r_{1m} \\ \vdots \ddots \vdots \ddots \vdots \\ r_{i1} \cdots r_{ij} \cdots r_{im} \\ \vdots \ddots \vdots \ddots \vdots \\ r_{n1} \cdots r_{nj} \cdots r_{nm} \end{pmatrix} \in \mathbb{R}^{n \times m} (1 \leq i \leq n, 1 \leq j \leq m)$$

If the j-th user has collected item α, $r_{\alpha i} = 1$; if not, $r_{\alpha i}$ is determined by ProbS. In other words, $r_{\alpha i}$ equals to the sum of resources transferring from those initial "resources" of items which the user has collected.

Obviously, all of those methods depend on the characters of bipartite network's topological structure. Popular items hold the trump card. Some unpopular items are unlikely to be recommended. Specifically, in ProbS method, their "resources" are even "0"s. Therefore, it is urgent to adopt a new approach which can improve the situation. In the next section, a novel concept will be introduced to resolve this dilemma.

### 3.3  Iterative Local Least Squares.

Least Squares method is a standard approach to the approximate solution of overdetermined system in mathematics, but it is also well attracted from other subjects. ILLS method is a kind of specialized forms which will be used to estimate missing value in a target user's rating vector where $r_{ij}$ equals 0 after using ProbS, one first chooses K nearest users using distance measure defined as following [20]:

$$Sim(u_i, u_j) = \frac{\sum_{\alpha \in O} a_{i\alpha} a_{j\alpha}}{\sqrt{k_i^2 + k_j^2}} \quad (2)$$

In addition, K is not fixed and those K users are regarded as coherent users to the target user. The missing values in coherent users' rating vector are filled with their respective row averages.

To explain how iterative local least squares works, we assume $\alpha$ is the lost rating value of the 1-th user for 1-th item. In fact, this value equals to w which is divided into the probe set (see section 4). Suppose the K neighbor users of 1-th user have been selected and are respective $t_1$-th user、$t_2$-th user…$t_{K-1}$-th user、$t_K$-th user. The submatrix consists of ratings given by 1-th user,, $t_1$-th... $t_K$-th user.

$$(r_1^T \quad r_{t_1}^T \quad \cdots \quad r_{t_{K-1}}^T \quad r_{t_K}^T) = \begin{pmatrix} \alpha & w^T \\ a & B \end{pmatrix}$$
$$= \begin{pmatrix} \alpha & w_1 & w_2 & \cdots & w_{K-1} & w_K \\ a_1 & b_{11} & b_{12} & \cdots & b_{1,K-1} & b_{1,K} \\ \vdots & \vdots & \vdots & \ddots & \vdots & \vdots \\ a_{n-1} & b_{n-1,1} & b_{n-1,2} & \cdots & b_{n-1,K} & b_{n-1,K} \end{pmatrix}$$

The least squares problem is then formulated as:

$$\min_{x} \left\| B^T x - w \right\| \quad (3)$$

Then, the missing α can be estimated as a linear combination of first values of K users

$$\alpha = a^T x = a^T (B^T)^{-1} w \qquad (4)$$

Let $x^* = (x_1, x_2 \cdots x_{n-1})$ denote the vector such that the square is minimized, we get

$$\alpha = x_1 a_1 + x_2 a_2 + \cdots + x_K a_K = \sum_{t=1}^{K} x_t a_t \qquad (5)$$

$$w \approx x_1 a_1 + x_2 a_2 + \cdots + x_K a_K = \sum_{t=1}^{K} x_t a_t \qquad (6)$$

During this process, a few elements of $r_{t_i}^T (1 \leq i \leq k)$ must be missed or equal 0. In this case, we have to fill those missing values in coherent with users' respective row averages. After imputing the estimate, the missing values can be replaced by such estimations

Before introducing ILLS, we need to present measures of performance. Normalized root mean squared error (NRMSE) [27] is a nice choice. Let SE be the set of missing values in R. Every missing rating has its true value v. While the estimate for this entry is $v^*$. Besides, the mean of those true values is represented by $\bar{v}$.

$$\bar{v} = \frac{1}{|SE|} \sum_{y_i \in SE} v \qquad (7)$$

The difference $|v - v^*|$ is the imputation error. Let μ and σ define as following:

$$\mu = \frac{1}{|SE|} \sum_{y_i \in SE} (v - v^*)^2 \qquad (8)$$

$$\sigma = \sqrt{\mu = \frac{1}{|SE|} \sum_{y_i \in SE} (v - \bar{v})^2} \qquad (9)$$

Then NRMSE is defined as

$$NRMSE = \frac{\sqrt{\mu}}{\sigma} \qquad (10)$$

NRMSE as an index used to measure the root's error of overdetermined equations. Because the overdetermined equation has no close-form solution, we just can find approximate solution. When one approximate solution makes residuals least, it will be set as the solution for this equation. NRMSE is a measure of the residual. And the lower RNMSE is, the better performance the proposed approach will get.

Another major problem is how to determine the value of K. In this paper we use one method to select nearest neighboring users. First, we fill those missing values in coherent users with their respective row average values. Second, for every value of K ranging from 1 to the total number of users in the dataset, it runs LLS once to estimate the values in probe dataset to determine the optimal value of K by NRMSE. Denote P as this optimal value.

After determining the value of K, in first iteration, we first use P to select neighboring users to run LLS to impute the missing values for each target user. Afterwards, at each iteration, ILLS uses their imputed results from last iteration to re-select the coherent users for every target user, using the same P [26]. The iteration process will stop when NRMSE converges at a certain low level.

## 4 Experimental results

### 4.1 Data Sets

We use three representative datasets to evaluate the performance of recommendation algorithms. The first one is *Movielens1M* data which can be download from the GroupLens site. The second one, *Netflix*, is a randomly selected subset of the huge dataset provided for the Netflix Prize. The last one, *RYM,* is obtained from http://RateYourMusic.com. Their basic statistical properties are shown in Table 1:

| Data sets | Users | Items | Links | Original Sparsity | Sparsity After Processing |
|---|---|---|---|---|---|
| *Movielens1M* | 943 | 1,682 | 10,000 | 0.0630 | 0.9969 |
| *Netflix* | 500 | 2,796 | 33,705 | 0.0241 | 0.9613 |
| *RYM* | 500 | 2,675 | 9,114 | 0.0068 | 0.5526 |

**Table 1.** Basic statistical properties of the three datasets.

### 4.2 Evaluating Indicators

An item is considered to be collected by a user only if the given rating is three or more (for *Movielens1M* and *Netflix*) or six or more (for *RYM*). To test a recommendation method on a dataset, we randomly remove 10% links (called probe set) and apply the algorithm to the remainder (training set) to produce a recommendation list for each user. We then employ three different metrics, to measure accuracy in recovery of these deleted links.

To evaluate the recommendation performance, we adopted four popular metrics: *AUC* [30], *Precision* [31], *Recall* [31], and *Diversity* [19]. AUC is the probability that the score of a link in test dataset is larger than a random link that does not exist. Precision is defined as the ratio of true positives in the top L recommendation items. If there are m items which are accurately predicted. Then $\Pr ecision = \frac{m}{L}$ . Recall shows the possibility that the items user may be potentially liked and recommended to the very user. It can be described as the ratio of truly positive recommended items to all the links in test set. Let $N_p$ represent the number of items which the target user likes in the probe dataset and $N_r$ represents the number of items that he/she likes in the recommendation list. $\text{Re}call = \frac{N_r}{N_P}$. One last metric is the diversity which characterizes how the recommendation method helps users enlarge their vision. It is defined as: $Diversity = \frac{2}{n(n-1)} \sum_{i \neq j}(1 - \frac{|I_R^i \cap I_R^j|}{L})$, where $I_R^i$ is the items recommended to user $i$.

### 4.3 Experimental Results

In the following, we show the experimental results for the ProbS and improved algorithms.

| Data Sets | Movielens1M | RYM |
|---|---|---|
| AUC | 0.563 | 0.549 |

**Table 2.** The AUC using ILLS without preprocessing ratings where K equals 30 percent of the number of users.

| Data Sets | ProbS AUC | ILLS AUC |
|---|---|---|
| Movielens1M | 0.905 | 0.905 |
| Netflix | 0.900 | 0.903 |
| RYM | 0.745 | 0.801 |

**Table 3.** AUC results of ProbS and ILLS in the three data sets.

As discussed above, it is obvious that using original ratings directly in ILLS method is not feasible due to the sparsity problem. So some measures must be taken. From table 2, after preprocessing, we can find ILLS is better than ProbS for all the three datasets respecting with AUC regardless of the recommendation length. That is because it becomes a very dense matrix after we use ProbS.

Fig.1 - Fig.5 show detailed experimental results. Fig. 1 shows how the convergence process of the iterations and the effect of K on it. It shows choosing 80% of the neighbors would give the best performance, and the iterative process quickly converge after about five iterations for *RYM*. Fig.2 - Fig.4 show the recommendation results of precision and recall for all the three data sets. We can see that ILLS almost keep the same accuracy with purely ProbS method. Similar results can also be found for the diversity result except *RYM*. It seems such length-based results might be resulted from the dense matrix which is already derived and used by ProbS.

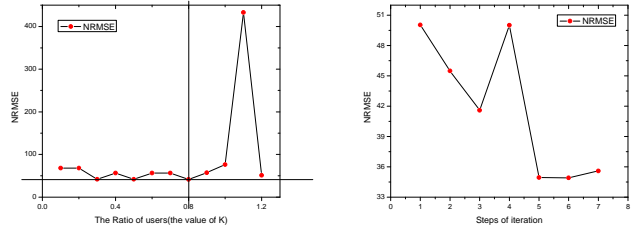

**Fig. 1.** NRMSE versus K for ILLS in *RYM* dataset with different ratios (left panel), and NRMSE versus the iterative steps (right panel) vs. iteration steps.

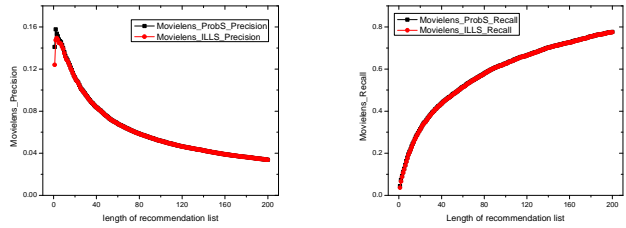

**Fig. 2.** Precision (left) and recall (left) vs. recommendation length for *Movielens1M*.

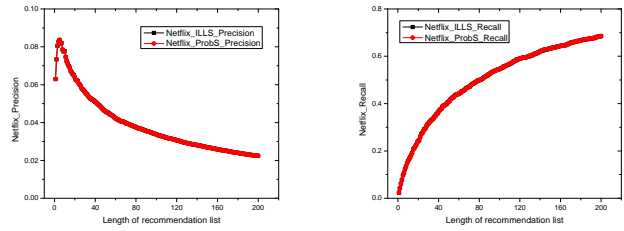

**Fig. 3.** Precision (left) and recall (left) vs. recommendation length for *Netflix*.

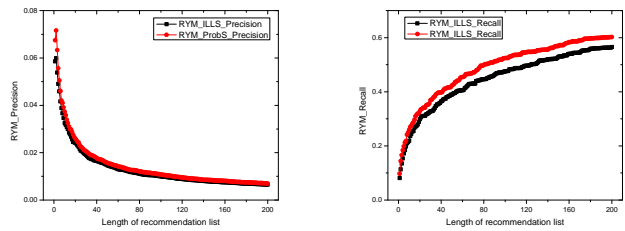

**Fig.** 4**.** Precision (left) and recall (left) vs. recommendation length for *RYM*.

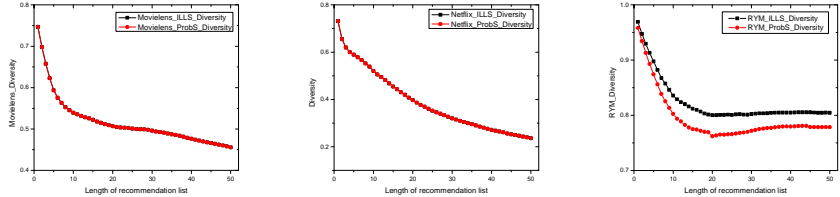

**Fig. 5.** Diversity vs. recommendation length for the three representative datasets.

## 5  Conclusions

In this paper, we proposed a two-step recommendation algorithm by iteratively increase the accuracy of prediction. Especially when facing sparse data, this approach promises to make recommendations more accurately than ProbS algorithm in three datasets, especially for the AUC results. While the sparse datasets result in bad results for many recommender methods, ILLS has offered a kind of newly possible way to provide suitable items for target users with little awareness of user behaviors. Another important respect is that the introducing of the new concept which makes us more conscious of relationship between the two subjects. This is just a start point and it is expected to adopt some more comprehensive sparsity-overcome methods, and eventually solve the cold-start problem.